\documentclass{article}
\usepackage[utf8]{inputenc}
\usepackage[square, numbers]{natbib}
\usepackage{parskip}
\usepackage[margin=2cm]{geometry}
\usepackage{graphicx}
\usepackage{epstopdf}
\usepackage{xcolor}
\usepackage{hyperref}
\usepackage{amsmath}
\usepackage{textcomp}

\title{Machine learning reveals complex behaviours in optically trapped particles}

\author{Isaac C. D. Lenton$^{1,*}$, Giovanni Volpe$^2$,
    Alexander B. Stilgoe$^1$, \\
    Timo A. Nieminen$^1$, Halina Rubinsztein-Dunlop$^1$
    \\[10pt]
    $^1$School of Mathematics and Physics, The University of Queensland, \\
    Brisbane, 4072, Australia \\
    $^2$Department of Physics, University of Gothenburg, SE-41296 Gothenburg, Sweden\\[10pt]
    $^*$uqilento@uq.edu.au}
\date{December 2019}

\begin{document}


\maketitle

\textbf{%
Since their invention in the 1980s \citep{Ashkin1986May}, optical tweezers have found a wide range of applications, from biophotonics and mechanobiology to microscopy and optomechanics \citep{Ashkin1987Mar, Svoboda1993Oct, Berut2012Mar, Jones2015Dec}.
Simulations of the motion of microscopic particles held by optical tweezers are often required to explore complex phenomena and to interpret experimental data \citep{Ashkin1992, Bustamante2005Jan, Seifert2012Nov, Butaite2019Mar}.
For the sake of computational efficiency, these simulations usually model the optical tweezers as an harmonic potential \citep{Volpe2013Mar, Bowman2013Jan}.
However, more physically-accurate optical-scattering models \citep{rohrbach2005stiffness, Nieminen2007Jul, Borghese2007, Nieminen2014Oct} are required to accurately model more onerous systems; this is especially true for optical traps generated with complex fields \citep{Stilgoe2011Dec, Sukhov2011Nov, Taylor2015Aug, McGloin2005Jan}. 
Although accurate, these models tend to be prohibitively slow for problems with more than one or two degrees of freedom (DoF) \citep{Bui2017Jul}, which has limited their broad adoption.
Here, we demonstrate that machine learning permits one to combine the speed of the harmonic model with the accuracy of optical-scattering models.
Specifically, we show that a neural network can be trained to rapidly and accurately predict the optical forces acting on a microscopic particle.
We demonstrate the utility of this approach on two phenomena that are prohibitively slow to accurately simulate otherwise: the escape dynamics of swelling microparticles in an optical trap, and the rotation rates of particles in a superposition of beams with opposite orbital angular momenta.
Thanks to its high speed and accuracy, this method can greatly enhance the range of phenomena that can be efficiently simulated and studied.
}

Since their invention \citep{Ashkin1986May}, optical tweezers have been employed to discover and study new phenomena, including molecular motors \citep{Svoboda1993Oct, Bustamante2005Jan}, thermodynamics of small systems \citep{Seifert2012Nov, Berut2012Mar}, and microscopic organisms \citep{Ashkin1987Mar, Zhang2008Apr}.
In the study of these systems, numerical simulations are frequently used both to plan and to interpret experiments.
The particle trajectory is often obtained from Brownian dynamics simulations, where it is simulated as a series of sequential Brownian and deterministic steps \citep{Volpe2013Mar, Bowman2013Jan}, which respectively account for the thermal motion and the deterministic optical forces acting on the particle.
Typically, the most computationally expensive part of the calculation is determining the optical forces; furthermore, due to the randomness introduced by the thermal motion, these forces need to be calculated sequentially.
For complex optical phenomena, multiple trajectories with a large number of time steps may be needed in order to have sufficient statistics.
For example, when the inertia contributes to the particle dynamics, such as for the motion of particles in vacuum, the particle trajectories need to be simulated with sub-nanosecond resolution, so that even the simulation of a 1-minute trajectory requires on the order of $10^{12}$ time steps \citep{Gieseler2018May}.

In simple situations, the optical trap can be approximated with a harmonic potential, so that the optical force can be easily estimated as $F_{\rm ot} = -k x$, where $k$ is the stiffness of the optical trap and $x$ is the displacement of the particle from the equilibrium position \citep{Jones2015Dec, Bowman2013Jan}.
However, there are many scenarios where this model is insufficient, for example, when modelling a particle escaping an optical trap \citep{Bui2015Sep} or the motion of a particle in a beam with orbital angular momentum \citep{Allen1992, Simpson2009Mar}.
While ad hoc phenomenological force-field models are often employed to simulate these phenomena \citep{Bui2013Apr, Phillips2014Apr}, it is necessary to employ semi-analytical methods such as the T-matrix method to obtain a physically accurate simulation \citep{rohrbach2005stiffness,  Borghese2007, Nieminen2007Jul, Nieminen2014Oct}.
However, these semi-analytical methods are prohibitively slow for running extensive dynamics simulations \citep{Bui2017Jul}.

An alternative approach is to use local interpolation, which involves sampling the force distribution at a number of discrete points on either a grid (structured interpolation) or at random locations (unstructured interpolation) and using an interpolation function (e.g., a linear or cubic polynomial) to estimate intermediate values \citep{press2007numerical}.
Unlike Brownian dynamics simulations, which require sequential force calculations, the force values required for interpolation can be calculated in parallel, significantly reducing run-time.
While interpolation works well for low-dimensional problems (1 to 3 DoF) \citep{press2007numerical}, it is difficult to implement efficiently for higher-dimensional problems because of runaway requirements on the memory storage and the lack of ready algorithms implemented in the major software packages.

In this Letter, we demonstrate an alternative method based on machine learning.
We show that a neural network (NN) can be trained to calculate optical forces with greater accuracy than phenomenological methods, in significantly less time than exact methods, and with less memory requirements than interpolation methods.
To demonstrate the power of this approach, we employ it to study two phenomena that would be prohibitively slow to accurately simulate otherwise: the escape dynamics of swelling micro-particles in an optical trap, and the hopping rates of particles in
a superposition of Laguerre-Gaussian beams.

\begin{figure*}
    \centering
    \includegraphics{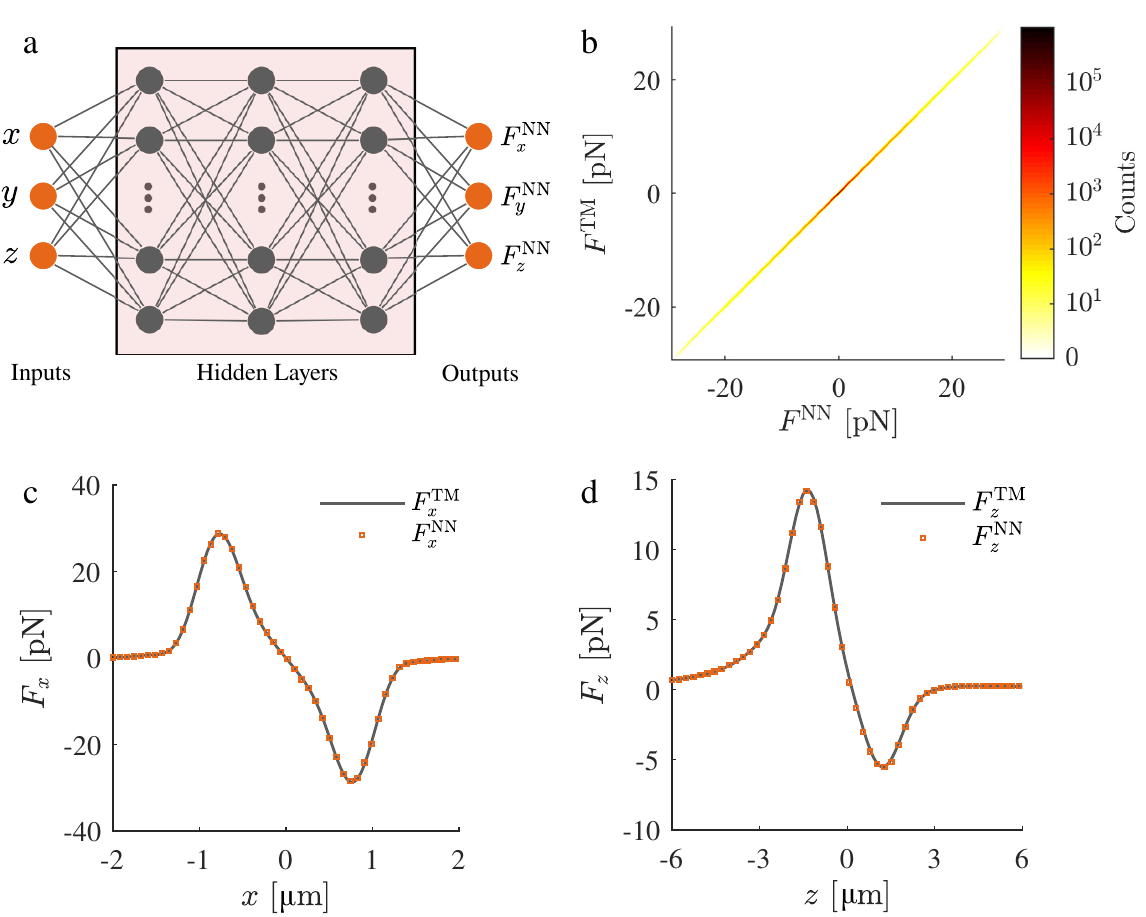}
    \caption{
    {\bf Estimation of optical forces using a neural network.}
    {\bf a} Illustration of a fully connected neural network with three inputs ($x, y, z$), three outputs ($f_x, f_y, f_z$), and three hidden layers.
    {\bf b} Log-density plot comparing neural network force estimates ($F^{\textrm{NN}}$) with the ground-truth forces calculated with the T-matrix method ($F^{\textrm{TM}}$) featuring an excellent agreement between the two methods for $10^6$ randomly sampled force vectors (viz. $3\times 10^6$ unique counts for the force components).
    {\bf c} Axial and {\bf d} radial force--position curves calculated using the neural network (orange symbols, $F^{\textrm{NN}}_x$ and $F^{\textrm{NN}}_z$) featuring an excellent agreement with those calculated with the T-matrix method (gray line, $F^{\textrm{TM}}_x$ and $F^{\textrm{TM}}_z$).
    }
    \label{fig:1}
\end{figure*}

While standard methods require explicit formulas to determine the optical forces starting from the physical parameters of the particle and trapping beam, in this work we employ machine-learning models that are trained to automatically determine the optimal rules to calculate the optical forces through a large series of known samples, i.e., pairs of physical parameters and corresponding optical forces (which can be obtained either from an experiment or an accurate numerical simulation).
Specifically, we employ NNs because they have been one of the most successful tools for machine learning in recent years \citep{LeCun2015May, Schmidhuber2015Jan}.
NNs have outperformed standard approaches to track particles \citep{Hannel2018Jun, Helgadottir2019Apr}, to enhance microscopy \citep{Rivenson2017Nov}, to predict optical scattering \citep{Peurifoy2018Jun}, and to calculate hydrodynamic forces \citep{Gibson2019Apr}.
Here, we employ dense fully-connected NNs, an example of which is illustrated in Fig.~\ref{fig:1}{\bf a}. 
These NNs consist of a series of fully-connected layers of neurons, where the output of each neuron is a nonlinear function of the weighted sum of the neuron's inputs.
The weights are iteratively adjusted (trained) until the NN learns to associate the correct optical force to each set of physical parameters.
To train the networks we used Keras with a Tensorflow backend \citep{chollet2015keras} (see Methods).

To evaluate the performance of this approach, we first explore the simple scenario of the optical forces acting on a spherical particle in an optical trap generated by a Gaussian optical beam.
The NN, shown in Fig.~\ref{fig:1}{\bf a}, consists of 3 layers of
256 neurons connected to three input and three output neurons.
The three inputs are the three-dimensional particle
position ($x$, $y$, $z$) and three outputs are the components of the optical force acting on the particle ($F^{\rm NN}_x, F^{\rm NN}_y, F^{\rm NN}_z$).
We train this NN using $10^6$ samples randomly distributed around the beam
focus, obtained from exact T-matrix simulations \citep{Nieminen2007Jul, Lenton2019Jul}.
Fig.~\ref{fig:1}{\bf b} compares the NN predicted forces with the exact
T-matrix forces for $10^6$ previously unseen samples, demonstrating very good agreement between the NN predictions and the T-matrix method.
Figs.~\ref{fig:1}{\bf c} and \ref{fig:1}{\bf d} further demonstrate the excellent agreement between the NN (orange symbols) and the exact T-matrix method (gray lines) both along the transverse $x$-direction (Fig.~\ref{fig:1}{\bf c}) and along the axial $z$-direction (Fig.~\ref{fig:1}{\bf d}), which extends well beyond the linear-force regime around the focal point.

\begin{figure*}
    \centering
    \includegraphics{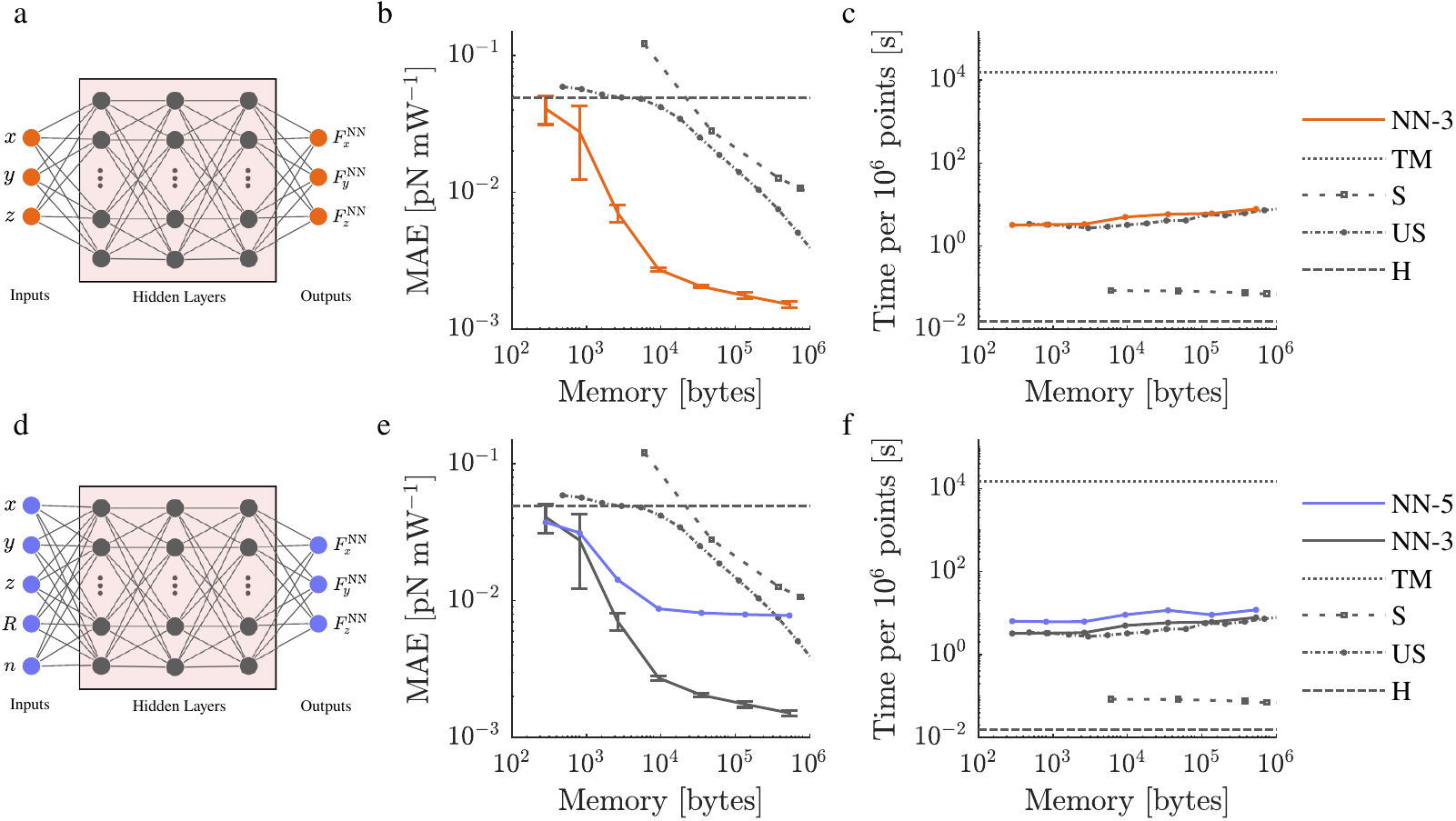}
    \caption{{\bf Improved performance of neural networks compared to other methods.}
    {\bf a} NN with 3 DoF (position $x$, $y$, $z$):
    {\bf b} as a function of their memory footprint, this NN (NN-3, orange solid line) outperforms in terms of mean absolute error (MAE) structured interpolation (S) and unstructured interpolation (US) as well as the harmonic model (H);
    {\bf c} furthermore, this NN (NN-3, orange solid line) is several orders of magnitude faster than the T-matrix (TM) and similarly fast as the unstructured interpolation (US), while the structured interpolation (S) and the harmonic approximation (H) are even faster.
    {\bf d} NN with 5 DoF (position $x$, $y$, $z$, particle radius $R$, refractive index $n$):
    {\bf e} its MAE and
    {\bf f} its speed are still comparable to those of the 3 DoF NN, while the other methods cannot be easily employed as discussed in the text; 
    for convenience of comparison, the gray lines in {\bf e} and {\bf f} report the results in {\bf b} and {\bf c}, respectively.
    Error bars in {\bf b} show the variation between 10 NNs trained on the same training data (see Methods for details).
    }
    \label{fig:2}
\end{figure*}

Fig.~\ref{fig:2} compares more quantitatively the NN and alternative standard methods in terms of both accuracy and speed as a function of their memory footprint.
We start by considering the 3 DoF NN discussed in Fig.~\ref{fig:1}.
By varying the number of neurons in the 3 intermediate layers, we directly
control the NN complexity and correspondingly the required memory.
This allows for a straightforward comparison between the NN memory
footprint and the memory needed to store the interpolation data
(see Methods).
The resulting mean absolute error (MAE) as a function of the memory footprint is shown by the orange line in Fig.~\ref{fig:2}{\bf b} (NN-3).
The NN accuracy is strongly dependent on the training sample distribution: for optimal training performance, the samples should be clustered around regions where the force varies rapidly.
Since in a Gaussian-beam optical trap the forces are largest and vary the most
around the beam focus, a convenient choice for the sample distribution is Gaussian-distributed samples around the focus (see Methods).
Similar considerations apply for the choice of points for unstructured interpolation (see Methods).
For all memory footprints, the NN greatly outperforms both structured interpolation (S, gray squares) and unstructured interpolation (US, gray circles).
This improved performance is due to the fact that, while unstructured interpolation requires storing all training samples indefinitely, NN only needs to see the training points during training and is able to encode the same information in a network using significantly
less memory.
For reference, we show also the MAE of the harmonic model (H, dashed line).
We can observe that, while for sparse grids (low memory footprint) both interpolation methods perform worse than the harmonic model, the NN already outperforms it even for the smaller memory footprints.

Fig.~\ref{fig:2}{\bf c} compares the performance of the various methods in terms of their computational speed on a standard laptop computer (3.4 GHz processor, 16 GB RAM, see Methods).
The exact T-matrix method is the slowest (TM, gray dotted line, $\approx10\,000\,{\rm s}$ to calculate $10^6$ samples), while the harmonic approximation is the fastest (H, gray dashed line, $\approx0.01\,{\rm s}$ to calculate $10^6$ samples). For both  these methods the memory footprint is fixed.
The NN (NN-3, orange line) and unstructured interpolation (US, gray circles) perform similarly ($\approx10\,{\rm s}$ to calculate $10^6$ samples) and about three orders of magnitude faster than the T-matrix method almost independently of the memory footprint.
Structured interpolation (S, gray squares) is faster than both these methods, but at the expense of a much lower accuracy (Fig.~\ref{fig:2}{\bf b}).

The results in Figs.~\ref{fig:2}{\bf b} and \ref{fig:2}{\bf c} show that the NN is better in terms of accuracy and similar in terms of speed to unstructured interpolation for systems with $\le 3$ DoF.
The NN gains a real edge when increasing the number of DoF. 
For example, we can consider a problem with 5 DoF, corresponding to the particle position, $x$, $y$ and $z$, its radius $R$, and its refractive index $n$.
In such case, structured interpolation becomes unfeasible because of memory constraints (a 5-dimensional grid with comparable resolution reaches quickly in the Terabytes). 
Similar memory-management problems emerge when considering unstructured interpolation algorithms.
A NN (Fig.~\ref{fig:2}{\bf d}) can still be trained with enough samples to learn the 5-dimensional dependence of the force field achieving good accuracy (Fig.~\ref{fig:2}{\bf e}) at a similar speed as in the 3-DoF case (Fig.~\ref{fig:2}{\bf f}).

\begin{figure*}
    \centering
    \includegraphics{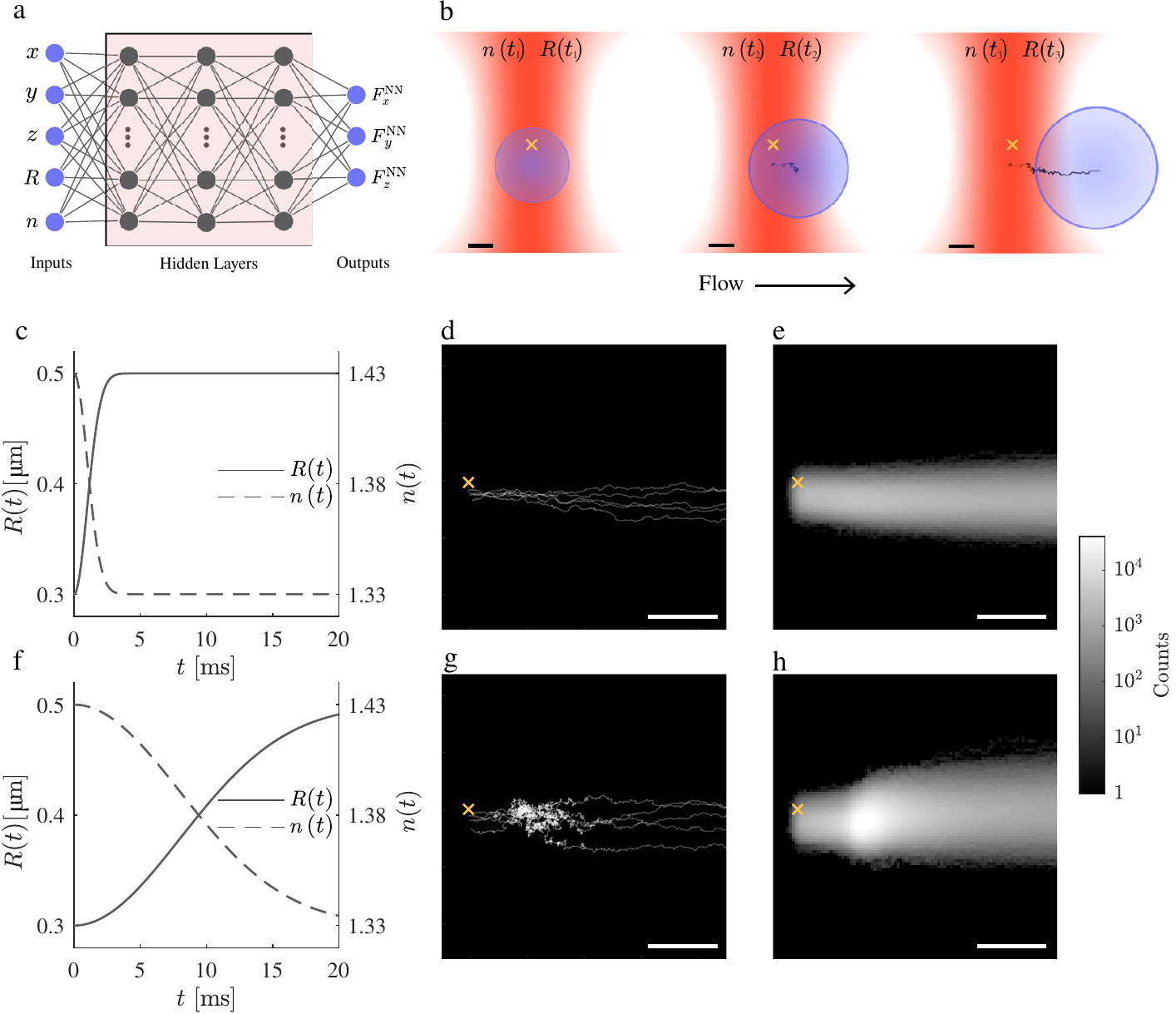}
    \caption{
    {\bf Swelling particle escape from an optical trap simulated with a 5-DoF neural network.}
    {\bf a} 5-DoF NN with inputs for particle position, $x$, $y$ and $z$, radius $R$ and refractive index $n$.
    {\bf b} A swelling particle is trapped in the presence of a flow; as the particles swells, its radius $R(t)$ increases and its refractive index $n(t)$ decreases, so that the restoring optical force decreases and, eventually, the particle escapes from the trap.
    The beam is propagating downwards and the trap centre is marked by the orange cross.
    {\bf c} Time dependence of the particle properties $R(t)$ (solid line) and $n(t)$ (dashed line) for a case in which the particle swells quickly.
    {\bf d} Example of 5 escape trajectories computed using the T-matrix method and {\bf e} $10^4$ trajectories simulated using the 5-DoF NN taking about the same computational time ($\approx 50\,{\rm s}$).
    {\bf f}-{\bf h} Corresponding results for a particle swelling more slowly.
    The scale bars in {\bf b}, {\bf d}, {\bf e}, {\bf g} and {\bf h} represent $0.2\,{\rm \mu m}$.
    }
    \label{fig:3}
\end{figure*}

We now turn our attention towards scenarios that are almost impossible
to accurately simulate without the NN approach.
First, in Fig.~\ref{fig:3}, we consider a particle whose size and refractive index gradually change while held in an optical trap.
This scenario is relevant to problems including
modelling hydrogel particles
swelling over time in optical tweezers or the study of
cells undergoing osmotic stresses.
More generally, this is an example of a problem where particles deform as they move through optical fields, such as the study of water-glycerol droplets \citep{Bae2019Feb} and of the deformation and growth of cells or microorganisms \citep{Norregaard2017Mar}.
Using the 5-DoF NN shown in Fig.~\ref{fig:3}{\bf a} (the same as that we employed in Figs.~\ref{fig:3}{\bf d}-{\bf f}), we simulate the escape trajectories of a swelling particle held in an optical trap in the presence of a fluid flow (see Methods).
As schematically shown in Fig.~\ref{fig:3}{\bf b}, a
swelling particle, whose radius increases and whose refractive
index decreases, is gradually pulled out of the trap when
the restoring optical force decreases.
This scenario is difficult to explore experimentally or computationally due to the large number of repetitions required to determine the statistics of the escape trajectories.
For a rapid swelling (Figs.~\ref{fig:3}{\bf c}-{\bf e}), the particle escapes the trap almost immediately after the flow is turned on.
For a slow swelling instead (Figs.~\ref{fig:3}{\bf f}-{\bf h}), the particle initially remains trapped at the
edge of the beam before eventually escaping.
Without the NN approach, it is practically impossible to collect enough statistics to clearly see how the escape trajectories change.
For example, using exact T-matrix calculations it takes us $\approx 50\,{\rm s}$ to simulate just 5 trajectories (Figs.~\ref{fig:3}{\bf d} and \ref{fig:3}{\bf g}), the same amount of time during which we can simulate $10^4$ trajectories with the NN (Figs.~\ref{fig:3}{\bf e} and \ref{fig:3}{\bf h}).
The latter amount of statistics are essential, for example, to determine the final particle distribution with a
high degree of accuracy.

\begin{figure*}
    \centering
    \includegraphics[width=\textwidth]{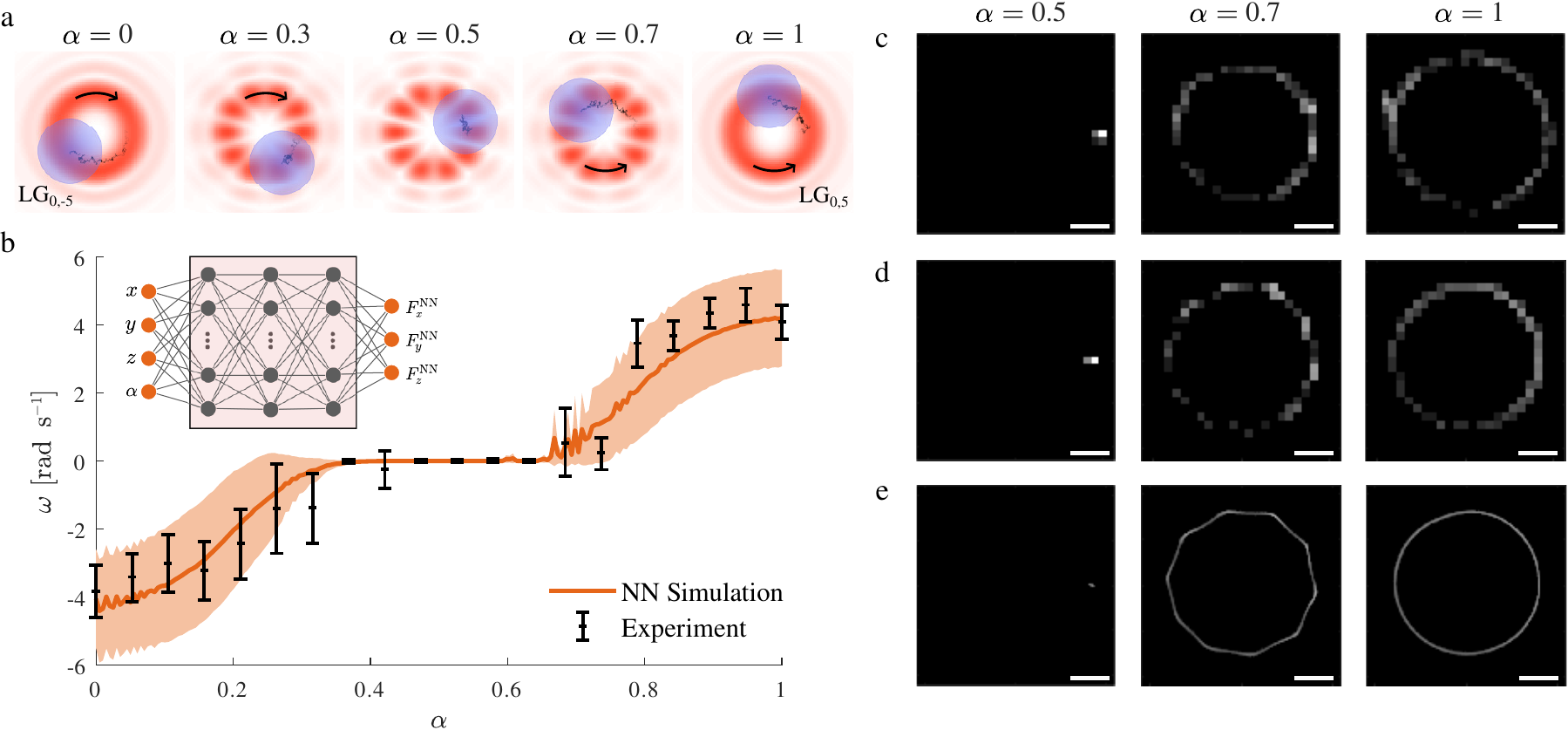}
    \caption{
    {\bf Dynamics of a particle in a superposition of Laguerre-Gaussian beams.}
    {\bf a} Schematic of an experiment with a particle held by a superposition of two Laguerre-Gaussian beams of order $\pm5$ with opposite orbital angular momentum, weighted by the  parameter $\alpha$ so that the total beam is ${\rm \alpha \, LG_{0, +5}} + (1-\alpha) \, {\rm LG_{0, -5}}$.
    {\bf b} Rotation rates obtained from experiments (black symbols) and from neural-network-based simulations (orange line). The error bars represent standard errors.
    Inset: schematic of the employed 4-DoF NN.
    {\bf c} 20-s experimental trajectory of a particle for various $\alpha$, leading to a confining potential ($\alpha = 0.5$), a washboard potential ($\alpha = 0.7$), and an inclined potential ($\alpha = 1$).
    {\bf d} Corresponding 20-s simulated trajectory for a single particle.
    {\bf e} Average of 100 simulated trajectories clearly showing the steady state behaviour in the various cases.
    The scale bars in {\bf c}-{\bf e} represent $1\,{\rm \mu m}$.
    }
    \label{fig:4}
\end{figure*}

The second scenario we consider involves the mixing of two beams with opposite orbital angular momentum (OAM) \citep{Padgett2011May}, which generate interesting interference patterns that can be used, for example, to design
ratchet-like micro-motors \citep{FrankeArnold2007Jul, Jesacher2004Aug}.
Fig.~\ref{fig:4}{\bf a} shows a schematic of the experiment.
A particle is placed in a mixture of two Laguerre-Gaussian beams of order $\pm5$
mixed into a single beam with a fixed power, so that the resulting beam is ${\rm \alpha \, LG_{0, +5}} + (1-\alpha) \, {\rm LG_{0, -5}}$, where $\alpha \in [0, 1]$.
Different values of $\alpha$ lead to different orbital rates $\omega$ for the optically trapped particle.
Fig.~\ref{fig:4}{\bf b} compares the results of an experiment (black symbols), where we trap a particle pushing it against the microscope slide using the laser beam (see details in Methods), with those we obtain employing a 4-DoF NN (shown in the inset of Fig.~\ref{fig:4}{\bf b}) to extensively simulate this system (orange line).
We obtain excellent agreement with the experimental results.
This strong agreement is further illustrated by the
time-averaged experimental (Fig.~\ref{fig:4}{\bf c})
and simulated (Fig.~\ref{fig:4}{\bf d}) trajectories, which
represent $\approx20\,{\rm s}$ of a single particle's trajectory, revealing three distinct behaviours:
When $\omega \approx 0$ ($\alpha = 0.5$), we have a series of confining potential wells resulting from the even mixture of two beams with opposite OAM \citep{Jesacher2004Aug}.
When $\omega$ is significantly larger than $0$ (e.g., $\alpha = 1$), we have a smooth inclined potential around the beam, where the particle slides at an approximately constant rate \citep{Padgett2011May, Allen1992}.
Finally, in intermediate regions (e.g., $\alpha = 0.7$), we obtain a washboard potential, where the particle is metastably trapped in local minima, but on average moves around the beam \citep{Dholakia2007Dec}.
The NN allows us to rapidly explore the effect of different
parameters (including height of the beam relative to the microscope slide,
beam power, and viscous drag) on the observed rotation rates and trajectories.
Further still, the NN allows us to simulate significantly more accurate
probability distributions: For example, Fig.~\ref{fig:4}{\bf e} shows
the average of 100 simulated trajectories, clearly displaying
the steady-state behaviour of this system.

Thus far, we have focused on how NNs can empower fast
and accurate simulations for designing, analysing and modelling experiments.
Another key advantage of the NN approach is its small memory footprint, which allows easy distribution of pre-trained NNs to calculate optical forces.
In this way, multiple users can take advantage of the initial training cost and perform sophisticated numerical simulations on readily available hardware.  This enables more collaborative workflows and the possibility
of integrating numerically accurate simulations into interactive
online demonstrations or teaching material.

In conclusion, we have shown NNs to be a valuable tool for simulating optical forces enabling longer, more accurate, and more memory-efficient simulations.
Compared to interpolation and other methods, NNs are easier to implement for problems with many DoF and they use significantly less memory while still performing with similar evaluation times to the most efficient interpolation techniques.
In particular, NNs can be used to supplement experiments, including to design trapping configurations or to analyse experimental results.
While here we considered only spherical particles, our method can be extended to particles with other shapes by adding rotations as inputs and torques as outputs.
Further investigation into different network architectures and training could further improve accuracy and decrease the number of points needed for training.
More broadly, the method described here is general enough to be applicable to a wide range of fields including simulations of optomechanics, optoelectronics or acoustics.
Although we focus on demonstrating this method
for simulating optically trapped particles, we
believe this technique will be of interest to the wider optics and photonics
community and can be used to model other complex optical processes.

\section*{Methods}

\subsection*{Training of Neural Networks}
NN are trained in Python using Keras (version 2.2.2) \citep{chollet2015keras} with TensorFlow backend (version 1.5.0).  An example Jupyter Notebook for the 3-DoF NN is provided in Supplementary Information.  Training each NN consists of three main parts: loading the data, setting up the NN model, and training the NN on the training data.

Training data for each NN is loaded from a data file containing exemplary inputs and outputs.  Values are read from the file, scaled to near unity, randomly shuffled and split into a validation and training set (with a 1/9 split).  For our NNs, training data consists of simulation data generated in Matlab (R2018a) using the Optical Tweezers Toolbox (OTT, version 1.5) \citep{Nieminen2007Jul, Lenton2019Jul}.  
Importantly, training data could have been generated with another process, for example, by position and force measurement in an experiment.

Setting up the NN model involves specifying the shape and parameters for each NN layer.  In this work we explored dense fully connected NNs with few intermediate layers.  Details about the number of layers and neurons per layer can be found in subsequent sections.  Intermediate layers use rectified linear unit (ReLU) activation functions.  The final layer uses a linear activation function.   Single precision values are used for NN weights for compatibility with available training hardware.

The NN is trained by optimising the NN weights to minimise a loss function. 
We use the Keras implementation of the Adam optimiser \citep{chollet2015keras} with default parameters and mean squared error for the loss function.  For training the NN, we use gradually increasing batch sizes.  The training set is used for training the NN and the validation set is used to evaluate the accuracy of the model at the end of each training epoch.  We train the NNs on NVIDIA Tesla V100 graphics cards.

\subsection*{3 Degrees of Freedom Neural Network}
These are NN with 3 inputs, 3 layers of hidden neurons and 3 outputs.  For the comparison presented in Figs.~\ref{fig:2}\textbf{b} and \ref{fig:2}\textbf{c}, we train 7 different NNs with 4, 8, 16, 32, 64, 128, and 256 neurons in each hidden layer.
Fig.~\ref{fig:1} uses 256 neurons for each hidden layer.  Training data is simulated with OTT for a spherical particle (radius 818 nm, refractive index 1.5) in a circularly polarised Gaussian beam (numerical aperture 1.02, medium refractive index 1.3, wavelength in medium 818 nm).
Training data consists of $10^6$ data points randomly distributed according to a normal distribution with 3.19 $\mu$m standard deviation and centred on the beam focus.  Error bars in Fig.~\ref{fig:2}\textbf{b} show average results of 10 NNs trained on the same data.

\subsection*{Mean Average Error}
Mean average error (MAE) is calculated according to
\begin{equation*}
    \text{MAE} = \frac{1}{3N_{\text{samples}}}
        \sum_i^{N_\text{samples}}
        \sum_{j=\{x, y, z\}}
        \left|
            F^{i,j}_{\text{TM}}
            - F^{i, j}_{\text{NN}}
        \right|
\end{equation*}
where $F_\text{TM}$ is the T-matrix optical force
calculated with OTT, $F_\text{NN}$ is the NN
estimate, and $N_{\text{samples}}$ is the
number of samples in the validation data set.
In Figs.~\ref{fig:1}\textbf{b},
\ref{fig:2}\textbf{b} and \ref{fig:2}\textbf{e},
the validation data set consists of $10^6$
additional locations not used in the training
of the NN.

\subsection*{Calculation of memory footprint}
Memory footprint depends on the method.
Optical tweezers harmonic models typically use between 1 and 9 values (i.e., between 8 and 72 bytes if using double precision values).
Structured interpolation stores values on an evenly spaced grid (i.e., a matrix), so that the required memory is directly related to the data type and the number of values stored (i.e., an interpolation grid with 
$100\times 100\times 100$ locations, 
each location containing 3 double precision values for $F_x$, $F_y$, and $F_z$,
has an approximate memory footprint of $24 \times 10^6$ bytes $\approx 24\,{\rm MB}$).
Unstructured interpolation requires storing the location and value at each grid point (i.e., for $10^6$ unstructured double precision position and force values, we need to store $48 \times 10^6$ bytes $\approx 48\,{\rm MB}$).  For the 3 DoF NN, the number of parameters (NN weights and biases) scales as
\begin{equation*}
    \text{Net}_3 = 2N^2 + 9N + 4
\end{equation*}
where $N$ is the number of neurons per hidden layer.  The memory footprint for the NN is approximately the number of parameters multiplied by the memory per parameter (4 bytes for single precision values).  The 5 DoF NN has an additional $2N$ parameters corresponding to the weights connecting to the two additional inputs, so that the total is 
\begin{equation*}
    \text{Net}_5 = 2 N^2 + 11 N + 4.
\end{equation*}

\subsection*{Sampling distribution}
For both unstructured interpolation and the NN approach, the choice of sampling locations affects the accuracy of the predicted forces.  With infinite samples, interpolation will exactly reproduce the target distribution.  With a suitable architecture and training, a NN is capable of emulating interpolation; consequently, the upper bound for the accuracy of a NN is related to the number and distribution of points used for training.  For the NN in this work, points are distributed around locations where the particle is likely to be and where the forces vary most rapidly.  For Gaussian beam NNs (3 DoF and 5 DoF), this corresponds to a Gaussian distribution centred at the beam focus.  For the 4 DoF NN, this corresponds to a toroidal shaped distribution; the shape of the toroid was determined from preliminary simulations which indicated where the particle was likely to be found.

\subsection*{Calculation of evaluation time}
To measure the evaluation time shown in
Figs.~\ref{fig:2}\textbf{c} and \ref{fig:2}\textbf{f},
we used a 3.4 GHz Intel Core i7-6700 personal computer with 16 GB RAM running implementations
of each algorithm in Matlab.
Matlab is chosen to allow simple comparison between performance of OTT (which is a Matlab package \citep{Nieminen2007Jul, Lenton2019Jul}) and the other methods.  Structured and unstructured interpolation use Matlab’s built-in implementations; and a harmonic model with 1 parameter per dimension is implemented.  Evaluation locations matching those used for the MAE comparison are used.  Methods are not explicitly parallelised.  Additional performance may be achievable using specialised implementations or explicit parallelisation.

\subsection*{5 Degrees of Freedom Neural Network}
NN with 5 inputs, 3 hidden layers and 3 outputs.  For Fig.~\ref{fig:2}, number of neurons in hidden layers is varied, similar to the 3 DoF NN.  For Fig.~\ref{fig:3}, each hidden layer has 256 neurons.  Training data is simulated with OTT for a spherical particle in a circularly polarised Gaussian beam (numerical aperture 1.02, wavelength in medium 800~nm, medium refractive index 1.33).
Training data consists of $10^7$ data points with: refractive index and radius values randomly sampled from a uniform distribution with the range 1.33--2.00 and 0.1--1.0~$\mu$m respectively; and positions randomly sampled from a normal distribution centred at the focus and standard deviation of 5~$\mu$m.

\subsection*{Escape trajectory simulation}
The particle refractive index and radius are given by
\begin{equation*}
    n(t, \sigma) = n_{\text{water}} + 0.1 e^{-t^2/2\sigma^2}
\end{equation*}
and
\begin{equation*}
    R(t, \sigma) = 0.5 - 0.2 e^{-t^2/2\sigma^2} \ \mu\text{m}
\end{equation*}
respectively, where $t$ is the simulation time, $\sigma$ is the decay (growth) rate of the refractive index (size) of the particle, and $n_{\rm water}$ is the refractive index of the medium.  The particle is initially trapped by the beam.  At 1~ms a flow is enabled causing the particle to move towards the edge of the trap and eventually escape.

\subsection*{Laguerre-Gaussian beam mixture experiment}
Particles are confined in two dimensions using mixtures of LG beams.
These beams carry orbital angular momentum which
is transferred to the particle.
Beams are created with a liquid crystal spatial light modulator (Meadowlark Optics, 512x512 HSP512L, high-speed SLM), illuminated by a 1064 nm fibre laser (YLR-10-1064-LP, 10 W, IPG Photonics).  The hologram is imaged onto the back-aperture of a 1.2 numerical aperture water immersion objective (Olympus UPlanSApo $60\times$) and focussed onto the sample.  The sample consists of 2~$\mu$m polystyrene spheres (refractive index $\approx$1.59) in water (refractive index $\approx$1.33) prepared between two microscope coverslips.  Particles are axially confined by pushing them against the microscope coverslip with the optical trapping beam.  A surfactant is mixed with the water to reduce the chance of particles sticking to the coverslip. 

\subsection*{4 Degrees of Freedom Neural Network}
NN with 4 inputs, 3 hidden layers with 128 neurons per layer, and 3 outputs.  Training data is simulated for a spherical particle (radius 1~$\mu$m, refractive index of particle 1.59) held in different mixtures of LG beams (refractive index of medium 1.33, wavelength in medium 800~nm).
In order to accurately model the experiment, training
data is generated using beams modelled on the experimentally realised beams (using the paraxial point-matching code from OTT and an approximation for the phase and intensity of the beam at the objective back aperture, numerical aperture 1.2).
The experimentally realised beams are approximately
LG+/-5 beams except for variations in the intensity
profile due to the incident illumination.
$10^7$ training points are generated.  Due to the computational expense in calculating different beam mixtures, only 21 different alpha values are present in the data set, evenly spaced from 0 to 1.
For each data point, alpha is randomly selected with a discrete uniform distribution, and position is uniformly randomly chosen from a toroidal volume with inner radius 0.5~$\mu$m and outer radius 2~$\mu$m and height 1~$\mu$m centred on the beam axis 0.5~$\mu$m from the focus.  The toroid corresponds to locations where the particle is trapped in the LG ring and allows multiple axial positions to be explored in simulations.

\subsection*{Laguerre-Gaussian beam mixture simulation}
The simulation is designed to model the experimental observations. The beam power is set to match the experiment rotation rate for a pure LG beam.  Axial position is chosen to produce a ring with the same radius as the experimentally observed ring.
Particle movement is constrained to prevent motion along the beam axis, emulating the effect of the microscope coverslip.  Particle motion is modelled using a finite difference simulation including both the deterministic optical forces and random Brownian motions.  The particle in the experiment was observed to occasionally stick to the slide; this is modelled with a frictional term introduced after including Brownian motion.

\section*{Funding}
This research was funded by the Australian Government through the Australian Research Council's Discovery Projects funding scheme (project DP180101002).
I.L. acknowledges support from the Australian Government RTP Scholarship.

\section*{Acknowledgements}
We would like to acknowledge the computing resources provided via the UQ SMP \verb|getafix| cluster.

\section*{Contributions}
This idea was conceived by G.V. and developed by G.V. and I.L..
I.L. generated the data sets used to train the NN.
G.V. and I.L. generated and trained the NNs.
I.L., T.N., H.R., and A.S. planned possible experiments.
I.L. conducted the experiment.
All authors contributed to the manuscript including planning figures.

\section*{Competing interests}
The authors declare no competing interests.

\bibliographystyle{unsrt}
\bibliography{main.bib}

\end{document}